\documentclass{article}

\usepackage[natbibapa]{apacite}

\usepackage[final]{neurips_2019}

\usepackage[utf8]{inputenc} 
\usepackage[T1]{fontenc}    
\usepackage{hyperref}       
\usepackage{url}            
\usepackage{booktabs}       
\usepackage{amsfonts}       
\usepackage{nicefrac}       
\usepackage{microtype}      
\usepackage{amsmath}
\usepackage{graphicx}

\usepackage{lipsum}
\newcommand\blfootnote[1]{%
  \begingroup
  \renewcommand\thefootnote{}\footnote{#1}%
  \addtocounter{footnote}{-1}%
  \endgroup
}

\title{Unsupervised predictive coding models may explain visual brain representation}

\author{%
  Marcio Fonseca\\
  Directorate of Technology and Innovation\\
  Chamber of Deputies\\
  Brasília, BR 70160-900\\
  \texttt{marcio.fonseca@camara.leg.br} \\
}

\begin{document}

\maketitle

\begin{abstract}
  Deep predictive coding networks are neuroscience-inspired unsupervised learning models that learn to predict future sensory states. We build upon the \emph{PredNet} implementation by \citet{lotter2016deep} to investigate if predictive coding representations are useful to predict brain activity in the visual cortex. We use representational similarity analysis (RSA) to compare PredNet representations to functional magnetic resonance imaging (fMRI) and magnetoencephalography (MEG) data from the Algonauts Project \citep{cichy2019algonauts}. In contrast to previous findings in the literature \citep{khaligh2014deep}, we report empirical data suggesting that unsupervised models trained to predict frames of videos may outperform supervised image classification baselines in terms of correlation to spatial (fMRI) data. Our best submission achieves an average noise normalized correlation score of $16.67\%$ and $27.67\%$ on the fMRI and MEG tracks of the Algonauts Challenge.
\end{abstract}

\section{Introduction}

Currently, convolutional neural networks trained on image recognition tasks are the best performing models to account for brain activity during visual object recognition \citep{schrimpf2018brain}. The performance of such supervised models on recent benchmarks led to the idea that supervised learning may be a requirement to explain visual cortex activity, especially in higher cortical areas \citep{khaligh2014deep}. In this work, we report experimental data suggesting that predictive coding models trained on unlabelled video data may outperform supervised baselines, yielding internal representations with higher of correlation to representation dissimilarity matrices (RDM) \citep{kriegeskorte2008} obtained from human fMRI and MEG data.\blfootnote{Code can be found at: \url{https://github.com/thefonseca/algonauts}}

This report summarizes our three main contributions. First, we find that a predictive coding model trained on videos captured with a car-mounted camera \citep{lotter2016deep} outperforms AlexNet \citep{krizhevsky2012imagenet} in terms of correlation to RDMs from human data. Moreover, as we further train the model on additional videos from the Moments in Time dataset \citep{monfort2018moments}, the model internal representations become more similar to brain activity. Second, we propose an end-to-end method to fine-tune predictive coding representations using joint supervision from frame prediction errors and IT dissimilarity scores. Our \emph{PredNet-IT} model improves the noise normalized correlation to human IT from $9.82\%$ to $15.93\%$ on the Algonauts Challenge test set \citep{cichy2019algonauts}. Lastly, we show that concatenating representations of PredNet-IT, and AlexNet improve our best MEG late interval correlation score from $17.15\%$ to $27.90\%$, which suggests these models capture complementary information relevant to visual recognition.

\section{Methods}
\label{methods}

\paragraph{Predictive coding networks}
We build on the \emph{PredNet} implementation by \citet{lotter2016deep}, which was shown to perform well on unsupervised learning tasks using video data. Inspired by the predictive coding theory \citep{friston2009predictive}, their model relies on the idea that to predict the next video frame, a model needs to capture latent structure that explains the image sequences. The PredNet architecture consists of recurrent convolutional layers \citep{xingjian2015convolutional} that propagate bottom-up prediction errors which are used by the upper-level layers to generate new predictions. For implementation details, please refer to the PredNet architecture description by \citet{lotter2016deep}.

\paragraph{Unsupervised training}
We evaluate predictive coding models trained on different quantities of unlabelled videos. The main idea is that the more data we use to train the model, the more "common sense" it should get about how events unfold in the world and, as a consequence, it should be better at disentangling latent explanatory factors. Using as starting point a PredNet pre-trained on the KITTI dataset \citep{geiger2013vision}, we further train the model with unlabelled videos from the Moments in Time dataset \citep{monfort2018moments}, a large-scale activity recognition dataset. Additionally, we report correlation scores for a PredNet with random weights and a version trained from scratch using just videos from the Moments in Time dataset.

The predictive coding model is trained in an unsupervised way to predict the next frames using a top-down generative model. The errors between predictions and the actual frames are propagated bottom-up to update the prior for new predictions. In terms of architecture, we follow the same hyperparameter settings used in the original PredNet implementation proposed by \citet{lotter2016deep}, with four modules (PredNet-4) consisting of $3\times3$ convolutional layers with $3$, $48$, $96$, and $192$ filters and input frames with dimensions $128\times160$. We also train a larger 5-layer model (PredNet-5) with $3$, $48$, $96$, $192$, and $192$ filters and input frames with a higher resolution of $256\times256$ pixels. The videos are subsampled at ten frames per second, and the network input is a sequence of ten frames for which the model generates ten frame predictions.

\paragraph{Feature extraction} 
Each layer $l$ and timestep $t$ of the PredNet model has representation units $R_l^{t}$, which are extracted as features to be compared to brain data. No further preprocessing or dimensionality reduction is performed. Since the Algonauts datasets consists of images and not videos, we repeat each image ten times to make the input compatible with the PredNet architecture.

Extracted features are then transformed to representational dissimilarity matrices (RDM) as described by \citet{kriegeskorte2008}. RDM serves as a common space to compare representations of different models and capture the dissimilarity (1 minus Pearson correlation) of internal representations generated for each pair of images in the dataset. To create RDMs from features, we use the Python implementation from the development kit provided by \citet{cichy2019algonauts}.

\paragraph{Evaluation}
The resulting RDMs for all model variants are compared to the RDMs from different brain regions, namely fMRI data from the early visual cortex (EVC) and the inferior temporal cortex (IT), and also MEG data for early and late stages of visual processing. The similarity of RDMs is computed in terms of Spearman's correlation, as defined by \citet{kriegeskorte2008} and implemented in the provided Python development kit. Evaluation is performed exclusively on the 92-images training set from the Algonauts Challenge, and the results are used to inform the choice of representations for submissions to the fMRI and MEG challenge tracks (78-images dataset). 

\begin{figure}
  \centering
  \includegraphics[width=.98\textwidth]{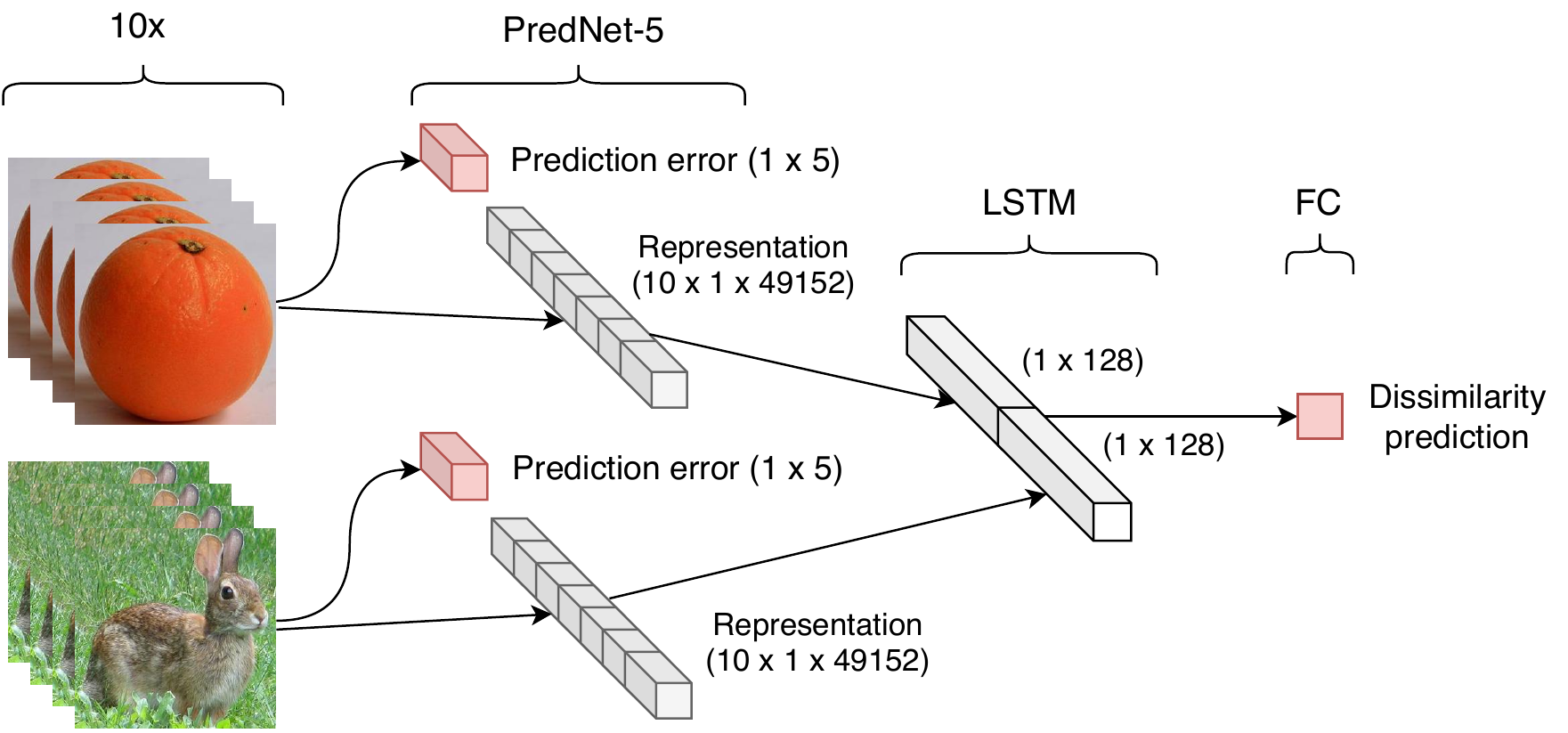}
  \caption{Fine-tuning predictive coding representations using RDM supervision. PredNet parameters are optimized to jointly minimize frame prediction error and dissimilarity error.}\label{fig:prednet-it}
\end{figure}

\paragraph{Fine-tuning representations}
Previous research suggests IT representation exhibits categorical clustering of concepts such as faces/non-faces and animate/inanimate entities \citep{khaligh2014deep}. Instead of assuming particular kinds of categories, we build a model to learn concept information implicitly from dissimilarity values. Specifically, to train our PredNet-IT model (Figure \ref{fig:prednet-it}), we sample 5,000 (training) and 500 (validation) image pairs from the 118-images training set and minimize a loss function $L$ combining the mean absolute frame prediction error $L_{PredNet}$ and the mean absolute error between predicted and ground truth dissimilarity given by the target IT RDM $L_{RDM}$ as follows:
\begin{equation}
    L = \alpha * L_{PredNet} + \beta * L_{RDM},
\end{equation}
where both $\alpha$ and $\beta$ weights are set to $1.0$. To infer dissimilarity scores, we apply an LSTM layer on top of the PredNet-5 (layer 5) representation. The resulting vectors are concatenated and passed to a fully connected layer that outputs the predicted scalar dissimilarity value.

\section{Results and Discussion}
\label{results}

\paragraph{Acquiring "common sense" improves correlation with brain data}
The noise normalized correlation scores on the 92-images training set (Table \ref{tab:92images}) shows that learning about how events unfold over time improves correlation scores. Notably, AlexNet (conv3 and conv4 layers) is outperformed by a PredNet pre-trained on 1 hour of videos from the KITTI dataset for both fMRI and MEG. Correlation to brain representation continues to improve as we train the PredNet-4 and PredNet-5 with up to 6 hours of videos from the Moments in Time dataset.

\begin{table}
  \caption{Correlation scores for different models and pre-training regimes. Correlation is evaluated on the \textbf{92-images training dataset} in terms of models RDM noise normalized squared correlation (Spearman) to fMRI and MEG RDMs. All PredNet features are from recurrent timestep 10 and all results are statistically significant ($p<0.05$).}\label{tab:92images}
  \centering
  \begin{tabular}{llcccc}
    \toprule
    & & \multicolumn{4}{c}{Noise Normalized $R^2$ (\%)} \\
    \cmidrule(r){3-6}
    & & \multicolumn{2}{c}{fMRI} & \multicolumn{2}{c}{MEG} \\
    \cmidrule(r){3-4}
    \cmidrule(r){5-6}
    Model Name     & Pre-training & EVC & IT & Early & Late \\
    \midrule
    AlexNet (conv3) & ImageNet & 17.85 & 7.14 & 5.29 & 13.31     \\
    AlexNet (conv4) & ImageNet & 13.98 & 10.66 & 2.72 & 16.68     \\
    \midrule
    PredNet-4 (layer 4) & random weights & 5.25 & 1.09 & 2.33 & 3.81    \\
    PredNet-4 (layer 4) & KITTI (1h) & 39.90   & 13.13 & 29.52 & 20.16  \\
    PredNet-4 (layer 4) & KITTI + Moments (4h) & 47.03 & \textbf{15.27} & 32.22 & 16.78  \\
    PredNet-5 (layer 3) & Moments (6h) & \textbf{51.48} & 14.52 & \textbf{40.35} & \textbf{22.89}  \\
    \bottomrule
  \end{tabular}
\end{table}

\paragraph{Fine-tuning improves correlation to IT}
Fine-tuning PredNet-5 with RDM supervision improves correlation to both fMRI and MEG data (except for late interval). Interestingly, fine-tuned PredNet-IT exhibits higher correlation to EVC and IT at layers 4 and 5, respectively as opposed to layer 3 in PredNet-5. Furthermore, there is a stronger concentration of high correlation scores for features from early recurrent timesteps, with different patterns for each layer. The best scores on the test set are $17.41\%$ (layer 3, timestep 6), $15.93\%$ (layer 4, timestep 10), $27.39\%$ (layer 4, timestep 8), and $1.95\%$ (layer 4, timestep 8) for EVC, IT, early and late intervals respectively (Table \ref{tab:78images}).

\paragraph{Unsupervised and supervised models capture complementary information}
The PredNet-IT model outperforms other supervised and unsupervised models across all data categories except late interval (MEG) (Table \ref{tab:78images}). To investigate the hypothesis that PredNet-IT captures complementary information, we generate new representations by merely concatenating PredNet-IT (layer 4, timestep 8) and AlexNet (conv4) activations. This combined representation results in a correlation score of $27.90\%$ for late interval, which is better than the AlexNet baseline ($22.93\%$) and all PredNet models. Further research is needed to check if this complementary information relates to "categorical clusters" similar to IT as found by \citet{khaligh2014deep}.

\begin{table}
  \caption{Correlation scores for submissions (\textbf{78-images test dataset}) in terms of models RDM noise normalized squared correlation (Spearman) to fMRI and MEG RDMs. PredNet-4 and PredNet-5 features are from recurrent timestep 10. See Section \ref{results} for details on timesteps for PredNet-IT.}\label{tab:78images}
  \label{sample-table}
  \centering
  \begin{tabular}{llcccc}
    \toprule
    & & \multicolumn{4}{c}{Noise Normalized $R^2$ (\%)} \\
    \cmidrule(r){3-6}
    & & \multicolumn{2}{c}{fMRI} & \multicolumn{2}{c}{MEG} \\
    \cmidrule(r){3-4}
    \cmidrule(r){5-6}
    Model Name & Pre-training & EVC & IT & Early & Late \\
    \midrule
    AlexNet (baseline) & ImageNet & 6.58 & 8.22 & 5.82 & 22.93     \\
    \midrule
    PredNet-4 (layer 4) & KITTI (1h) & 10.55 & 8.55 & 3.66 & 2.41  \\
    PredNet-4 (layer 4) & KITTI+Moments (4h) & 17.40 & 5.77 & 21.19 & 17.15  \\
    PredNet-5 (layer 3) & Moments (6h) & 17.00 & 9.82 & 14.83 & 6.97  \\
    PredNet-IT (layers 3, 4) & Moments (6h) & \textbf{17.41} & \textbf{15.93} & \textbf{27.39} & 1.95 \\
    PredNet-IT+AlexNet & Moments (6h)+ImageNet & 4.01 & 6.67 & 0.83 & \textbf{27.90} \\
    \bottomrule
  \end{tabular}
\end{table}

\paragraph{Final remarks and future directions}
\label{discussion}
Our results suggest that not only supervised but also unsupervised models may explain visual brain data. While our results are still far from the noise ceilings, there is still much room for improvement by scaling the model architecture and training data (AlexNet has almost ten times more parameters than the PredNet-4 model). An interesting future investigation could combine features from different layers to assess if predictive coding models can fully explain higher cortical areas and exhibit visuo-semantic representations.

\bibliographystyle{apacite}

\bibliography{references}
\small

\end{document}